\documentclass[twocolumn,preprintnumbers,amsmath,amssymb]{revtex4}
\usepackage{graphicx}
\usepackage{dcolumn}
\usepackage{bm}
\begin{document}
\title{Persistent spin current and entanglement in the anisotropic spin ring}
\author{Zi-Xiang Hu, You-Quan Li}
\address{Zhejiang Institute of Modern Physics, Zhejiang University,
Zhejiang, Hangzhou 310027, P. R. China}
\date{\today}
\begin{abstract}
We investigate the ground state persistent spin current and the
pair entanglement in one-dimensional antiferromagnetic anisotropic
Heisenberg ring with twisted boundary conditions. Solving Bethe
ansatz equations numerically, we calculate the dependence of the
ground state energy on the total magnetic flux through the ring,
and the resulting persistent current. Motivated by recent
development of quantum entanglement theory, we study the
properties of the ground state concurrence under the influence of
the flux through the anisotropic Heisenberg ring. We also include
an external magnetic field and discuss the properties of the
persistent current and the concurrence in the presence of the
magnetic field.
\end{abstract}
\maketitle

\section{introduction}

Transport properties of strongly correlated systems have attracted
great theoretical and experimental interest for more than two
decades. In particular, low-dimensional systems show significant
deviations in transport properties from the usual Fermi-liquid
quasiparticle description. It has been revealed that electron
correlation and topology play important roles in one-dimensional
systems. Recently, the study of transport properties in integrable
models has been an active field of research. The two most studied
models are the hubbard model and the spinless fermion(or,
equivalently, spin-1/2 Heisenberg chain) model. Several
experiments~\cite{LPLevy,VChandrasekhar,DMailly,BReulet,EMQJariwala,WRabaud}
have observed persistent currents in mesoscopic metallic and
semiconducting rings pierced by a magnetic flux. These studies
have led to many theoretical investigations focusing on the
interplay of electron-electron interaction and disorder in such
systems. The persistent current in a ferromagnetic Heisenberg ring
has been studied in crown-shaped magnetic
field~\cite{FlorianSchutz}, which can also be driven by an
inhomogeneous electric fields~\cite{ZLCao} due to the
Aharonov-Casher effect~\cite{YAharonov}. Based on a spin-wave
approach, the spin current in an antiferromagnetic Heisenberg ring
with integer spin in an inhomogeneous magnetic field has been
investigated very recently~\cite{F. Schutz}.

In the first part of this paper we study the ground state
persistent current of a spin-1/2 anisotropic Heisenberg spin ring
pierced by a flux. By solving Bethe ansatz equations numerically,
we find that the increasing anisotropy reduces the amplitude of
the persistent current. This phenomenon has also been noticed by
G.Bouzerar~\cite{Bouzerar} using the Lanczos method. We also study
the influence of an external magnetic field on the ground-state
persistent current, and find that the increasing magnetic field
reduces the amplitude of the persistent current. When the magnetic
field is larger than a certain value ($h=3.0$ for $\Delta=0.5$,
when the spins are fully polarized), the persistent current
vanishes.

Quantum entanglement, as exemplified in the singlet state of two
spin-1/2 particles
$(1/\sqrt{2})(|\uparrow\downarrow>-|\downarrow\uparrow>)$, is a
correlation between two quantum subsystems. It bears some
resemblance to classical correlation, but differs in important
aspects, highlighted by the observation of the violation of Bell's
inequalities in entanglement systems. For a special case of three
binary quantum objects(three qubits), a quantitative extension of
these inequalities has been proposed in terms of a measure of the
entanglement called the ``concurrence", which takes values between
zero and one. Several theoretical
studies~\cite{CHBennett,SHill,wootters} have pointed out, for
example, that the square of the concurrence between qubits A and
B, plus the square of the concurrence between A and C, cannot
exceed unity~\cite{VCoffman}. Meanwhile, it has been found
experimentally that entanglement is crucial to describe magnetic
behavior in a quantum spin system~\cite{Ghose}. Therefore, we hope
that the study of the entanglement in Heisenberg models will
enhance our understanding of the quantum features of magnetic
systems.

 We may regard a spin-1/2 chain as a collection of interacting
qubits. This connection has motivated us to carry out the second
part of the study to investigate the entanglement in anisotropic
spin chains. There have already been several studies on the
entanglement in spin
chains~\cite{MCArnesen,DGunlycke,wangPRA,GLagmago,YangSun,LFSantos,TJOsborne}.
In particular, recent
efforts~\cite{KMOConnor,VBuzek,sjgu2005,ClareDunning} have been
made to understand the quantum entanglement in the ground states
of some many-body models, in the hope that the study of
entanglement can provide a new insight into the quantum phase
transition~\cite{AOsterloh,GVidal} in these systems. Motivated by
a recent work about the entanglement generation in persistent
current qubits~\cite{JFRalph}. To this end, we study the
properties of the ground state concurrence and its connections
with persistent current. Our numerical results show that the
concurrence is a periodic function of $\phi$. The locations of the
minimum concurrence correspond to those of the maximum persistent
current where an energy level crossing occurs. In the presence of
the external magnetic field, the concurrence increases with the
magnetic field, though the minima become less sharp as the
magnetic field increases.
\section{the model and its secular equation}

The Hamiltonian of an anisotropic Heisenberg ring with twisted
boundary conditions reads:

\begin{equation}\label{hamiltonian}
H =  - J\sum\limits_{l = 1}^N {\left[\frac{1}{2}(S_l^ +  S_{l +
1}^ - e^{ - i\phi /N}  + H.c.) + \Delta S_l^z S_{l + 1}^z \right]}
\end{equation}
where $N$ is the number of sites. The total magnetic flux, or the
Aharonov-Bohm flux, through the ring is $\Phi$, and
$\phi=2\pi\Phi/\Phi_0$ where $\Phi_0=hc/e$ is the flux quantum.
$S_l^+$, $S_l^-$, and $S_l^z$ are spin-1/2 operators at site $l$.
 By using Wigner-Jordan transformation~\cite{Jordan}, the Hamiltonian Eq.~(\ref{hamiltonian}) can be
mapped to a spinless fermion model:
\begin{eqnarray}
H =  - \frac{J}{2}\sum\limits_{ i } ({C_i^ +  } C_{i+1} e^{-i\phi/N}+{C_{i+1}^ +  } C_{i} e^{i\phi/N})\nonumber\\
 + V\sum\limits_i {(\hat n_i  - \frac{1}{2})(\hat n_{i + 1} -
 \frac{1}{2})},
\end{eqnarray}
where $J/2$ can be interpreted as the hopping integral and
$V=J\Delta$ as the nearest-neighbor Coulomb repulsion. The
spinless fermionic operators $c_i^+$ and $c_i$ obey the
anticommutation relations, and $\hat n_i=c_i^+ c_i$ is the local
number operator. It is well known that the ground state of this
model has different phases: a metallic phase when $0<\Delta< 1$
and an insulating phase when $\Delta>1$, where $\Delta$ is the
anisotropy, or interaction. The former phase is gapless while the
latter gapful.

As in Ref.~\cite{Li}, the flux can be gauged out of the
Hamiltonian Eq.~(\ref{hamiltonian}), so that solving the
Schr\"odinger equation in the presence of a magnetic flux with a
periodic boundary condition is equivalent to that in the absence
of the flux but with a twisted periodic condition, namely,
\begin{equation}\label{TBC}
\psi (x_1 , \cdots x_i  + L, \cdots ) = \exp (i\phi)\psi (x_1 ,
\cdots x_i , \cdots ).
\end{equation}
We emphasize that the characteristic length $L$ of the
circumference is a mesoscopic scale so that inelastic scattering
does not occur. This justifies the applicability of the Bethe
ansatz approach for the present problem, because the ansatz
embodies nondiffractive scattering.

From the standard quantum inverse scattering methods
(QISM)~\cite{HABethe,takahashi}, the diagonalization of
Hamiltonian Eq.~(\ref{hamiltonian}) can be transformed into
solving the following Bethe ansatz equations:
\begin{equation}\label{logh}
\left( {\frac{{\sinh \frac{\gamma }{2}(x_j  + i)}}{{\sinh
\frac{\gamma }{2}(x_j  - i)}}} \right)^N  = e^{ - i\phi }
\prod\limits_{l \ne j}^M {\frac{{\sinh \frac{\gamma }{2}(x_j  -
x_l  + 2i)}}{{\sinh \frac{\gamma }{2}(x_j  - x_l - 2i)}}},
\end{equation}
where $\gamma$ satisfies $\Delta=\cos(\gamma)$.

 Taking the logarithm of Eq.~(\ref{logh}), we obtain
\begin{equation}\label{BAE}
N\theta _1 (x_j ,\gamma ) = 2\pi I_j  + \phi  + \sum\limits_{l =
1}^M {\theta _2 (x_j  - x_l ,\gamma )},
\end{equation}
where
$$
\theta _n (x,\gamma ) = 2\tan ^{ - 1} [\frac{{\tanh (\gamma
x/2)}}{{\tan (n\gamma /2)}}].
$$
The ground state is described by a symmetrical sequence of quantum
numbers around zero. They are
$$
I_j  = \{  - \frac{{M - 1}}{2}, - \frac{{M - 3}}{2},...,\frac{{M -
1}}{2}\}.
$$
The energy of the ground state can be expressed as:
\begin{equation}
E = E_0  - J\sum\limits_{j = 1}^M {\frac{{\sin ^2 (\gamma
)}}{\cosh (\gamma x_j )-\cos (\gamma )}}.
\end{equation}
In the thermodynamics limit, it can be written in an integral
form,

\begin{eqnarray}
E = E_0  - JN\sin ^2 \gamma \int {{{\rho _N (x,\phi )dx} \over {\cosh (\gamma x) - \cos \gamma }}} \nonumber\\
 \times
\sum\limits_{m =  - \infty }^\infty  {\exp \{ im[Np_N (x) - \phi
]\} },
\end{eqnarray}
where $ p_N (x_i ) = (2\pi I_i  + \phi )/N $ and $E_0=J\Delta
N/4$.

At $T=0$, the equilibrium persistent current can be given by
$I(\phi)=-(e/\hbar)\partial E(\phi)/\partial \phi$, where
$E(\phi)$ is the ground-state energy of Eq.~(\ref{hamiltonian}),
as a function of the boundary condition $\phi$ [Eq.~(\ref{TBC})].

\section{persistent spin current}
\begin{figure}
\includegraphics[width=7cm]{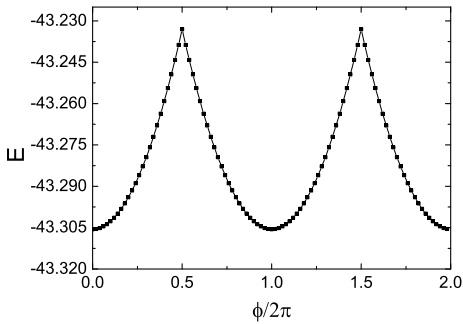} \caption{\label{Figure ground}
The ground state energy of anisotropic Heisenberg spin ring as a
function of flux when $\Delta=0.5403$.}
\end{figure}

\begin{figure}
\includegraphics[width=7cm]{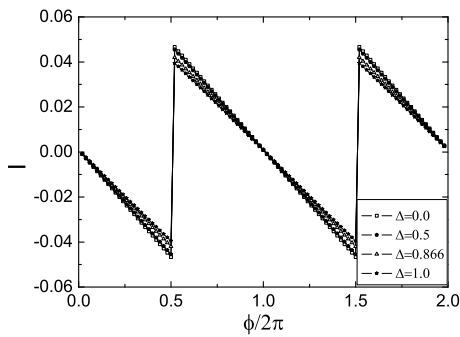} \caption{\label{Figure current}
The persistent current in anisotropic Heisenberg spin ring versus
flux for different anisotropic parameters
$\Delta=0,1/2,\sqrt{3}/2,1$. }
\end{figure}

To obtain the ground state of the anisotropic Heisenberg model,
which is a spin singlet state $(N=2M)$, we solve the Bethe ansatz
equations Eq.~(\ref{BAE}) at zero temperature and calculate the
dependence of the ground-state energy on flux. From the
Eq.~(\ref{BAE}), we find that when $\phi<\pi$, the quantum number
sequence of the lowest energy state should be the same as the
ground state when $\phi=0$. But if $\pi<\phi<3\pi$, the quantum
number sequence of the lowest energy state should be $ I_j = \{ -
\frac{{M + 1}}{2}, - \frac{{M - 1}}{2},...,\frac{{M - 3}}{2}\}$.
Likewise, for $3\pi<\phi<5\pi$, it should be $ I_j  = \{ -
\frac{{M + 3}}{2}, - \frac{{M + 1}}{2},...,\frac{{M - 5}}{2}\}$.
In other words, when $\phi$ reaches $(2n+1)\pi$, there is an
energy level crossing at the lowest energy level. As the flux
increased, the original ground-state energy increases while
excited energy decreases. They are the same when $\phi$ equals
$(2n+1)\pi$. Here we solve a system with 42 spins, and obtain the
ground state energy as in Fig.~(\ref{Figure ground}).
We can find that the ground state energy is a periodic function
with respect to the flux, with a  periodicity of $2\pi$. Using
$I(\phi)=-(e/\hbar)\partial E(\phi)/\partial \phi$, we can easily
calculate the persistent current as shown in Fig.~(\ref{Figure
current}) for different anisotropy. $I(\phi)$ has a sawtooth-like
shape, and is also a periodic function of flux with a periodicity
of $2\pi$. There is a discontinuity where the energy level
crossing occurs. The existence of a  discontinuity at finite
$\Delta$ comes from the fact that the translational invariance is
persevered in the presence of interaction ($V=J\Delta\neq0$). Li
and Ma \cite{Li} considered the current of an isotropic Heisenberg
ring (i.e. $\Delta=1$) using the Bethe ansatz method, and also
found the spin current as a linear function of the flux. In
Fig.~(\ref{Figure current}), we draw the persistent current with
different anisotropic parameters $\Delta$. The current is always a
periodic function of the AB flux, and the increasing anisotropic
parameter $\Delta$ reduces the amplitude of the persistent
current. In the regime of $0<\Delta<1$, the amplitude change is
weak, while is was found to decrease more rapidly  for $\Delta>1$
~\cite{Bouzerar}.

\begin{figure}
\includegraphics[width=7cm]{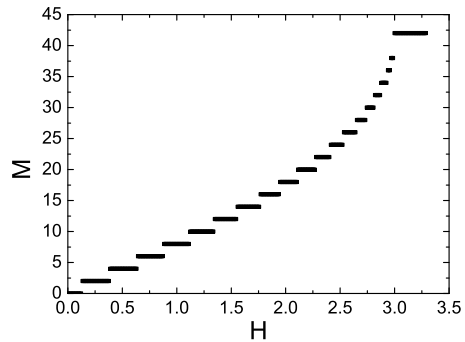} \caption{\label{Figure M-H}
The magnetization M of of the lowest energy state as a function of
external magnetic field when $\Delta=0.5$.}
\end{figure}

In the presence of external magnetic field, the hamiltonian
changes into $H=H_0+h\sum{\overrightarrow{S_z}}$, in which $H_0$
is the origin hamiltonian [Eq.~(\ref{hamiltonian})] without the
magnetic field. The magnetization
$\sum{\overrightarrow{S_z}}=\frac{1}{2}(N-2M)$,
$H=H_0+\frac{h}{2}(N-2M)$. For $h=0$, the ground state of this
system is spin singlet(N=2M), while for $h\neq 0$, the ground
state is no longer spin singlet, the magnetic field may flip some
spins. For example, when $h=0.2$, the lowest energy state of the
system with 42 spins is $M=20$ rather than $M=21$. In
Fig.~(\ref{Figure M-H}), we plot the dependence of the
ground-state magnetization as a function of the external field.
\begin{figure}
\includegraphics[width=7cm]{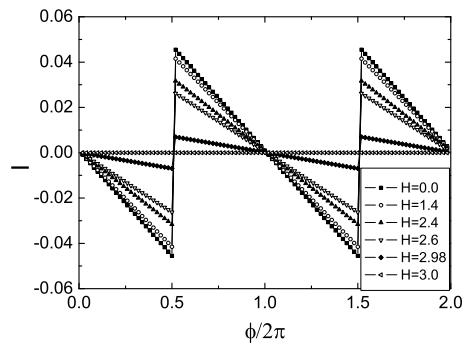} \caption{\label{Figure I-H}
The persistent current as a function of the flux under the
different external magnetic fields when $\Delta=0.5$.}
\end{figure}
We can then calculate the spin current. In Fig.~(\ref{Figure
I-H}), we plot the persistent current at different magnetic fields
for anisotropy $\Delta=0.5$. We find that the shape of the current
as a function of flux is unchanged, but the amplitude of the spin
current decreases when the external magnetic field increases. And
when $h$ is small the amplitude decreasing velocity is slower than
it when h is much larger. After $h\geq3.0$, the persistent current
decreases to zero because all the spins are polarized by the
external magnetic field.

\section{the pair entanglement}
\begin{figure}
\includegraphics[width=7cm]{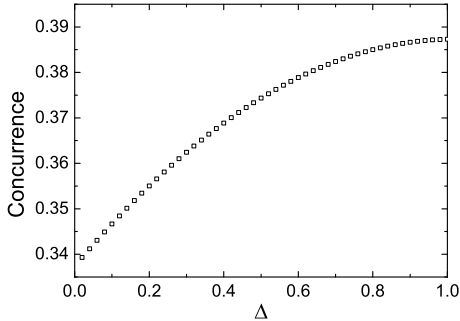} \caption{\label{Figure currentdelta}
The concurrence of XXZ spin ring as a function of $\Delta$ for a
system with 42 spins.}
\end{figure}

We now turn to the calculation of the ground state concurrence $C$
as a function of the total magnetic flux $\phi$. Because the
hamiltonian is invariant under translation, the entanglement
between any two nearest neighboring sites is independent of site
index. When $\Delta\neq1$, Eq.~(\ref{hamiltonian}) becomes
q-deformed SU(2) algebra with $\Delta=(q+q^{-1})/2$. Together with
the $Z^2$ symmetry, we have $[H,S^z]=0$, which simplifies the
reduced density matrix $\rho_{l(l+1)}$ of two neighbor sites as
$$
\rho _{l(l + 1)}  = \left( {\begin{array}{*{20}c}
   {u^ +  } & 0 & 0 & 0  \\
   0 & {w_1 } & z & 0  \\
   0 & {z^* } & {w_2 } & 0  \\
   0 & 0 & 0 & {u^ -  }  \\
\end{array}} \right)
$$
in the standard basis
$|\uparrow\uparrow>,|\uparrow\downarrow>,|\downarrow\uparrow>$,
and $|\downarrow\downarrow>$. The energy of a single pair in the
system is $E/N=Tr[\rho_{l(l+1)}H_l]$, where $H_l$ is the part of
the Hamiltonian between site $l$ and $l+1$, due to the
translational invariance. From the definition of entanglement, we
can easily find that the concurrence of anisotropic Heisenberg
ring can be calculated as \cite{wang2002}:
\begin{equation}\label{concurrence}
C = \frac{1}{2}\max (0,|E_{gs}/N - \Delta G_{l(l + 1)}^{zz} | -
G_{l(l + 1)}^{zz}  - 1),
\end{equation}
where $G_{l(l + 1)}^{zz}$ is the two-site correlation function
defined by $G_{l(l+1)}=Tr[\exp(-\beta H)\sigma_{l z}\sigma_{l+1
z}]/Z=-\frac{2}{N\beta}\frac{\partial lnZ}{\partial \Delta}$,
where $\beta=1/T$, and $Z$ is the partition function. $E_{gs}$ is
the ground-state energy. In order to obtain the concurrence, we
calculate the correlation first. From the definition of the
two-site correlation function, we know that once the
$\Delta$-dependent eigenenergy $E(\Delta)$ is obtained, the
correlation function is simply the first derivative of
$E(\Delta)/N$ with respect to $\Delta$. Then we obtain the
concurrence as a function of $\Delta$ as plotted in
Fig.~(\ref{Figure currentdelta}).
This result was also noticed in the Gu {\it et al} 's work
~\cite{Gu}. They obtained an approximative function around the
critical point $\Delta=1$ as $C_0-C_1(\Delta-1)^2$ where
$C_0=2\ln2-1\simeq 0.386$,
$C_1=2\ln2-\frac{1}{2}-\frac{2}{\pi}-\frac{2}{\pi^2}\simeq0.047$.
The largest value of concurrence at $\Delta=1$ is the result of
competition between quantum fluctuation and ordering. This seems
to be independent of the system size: five qubits were considered
in Ref.~\cite{wang}, while 1280 sites were considered in
Ref.~\cite{Gu}.

\begin{figure}
\includegraphics[width=7cm]{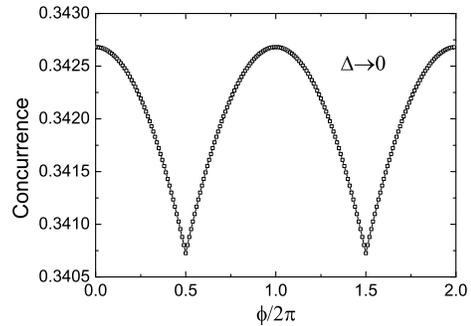} \caption{\label{Figure concurrence}
The concurrence as a function of flux when $\Delta\rightarrow 0$.}
\end{figure}

\begin{figure}
\includegraphics[width=7cm]{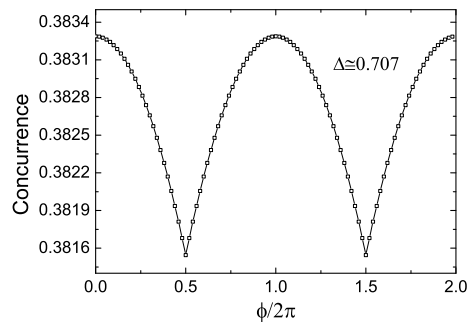} \caption{\label{Figure concurrence2}
The concurrence as a function of flux when $\Delta \simeq 0.707$.}
\end{figure}

 Further more, we study the influence of the magnetic flux.
 The concurrence as a function of
 the flux is plotted in Fig.~(\ref{Figure concurrence}) and in Fig.~(\ref{Figure concurrence2})
 for two different anisotropics $\Delta$.
We find that the ground-state concurrence is also a periodic
function of the flux. The locations of maximum concurrence are
those of minimum energy. When the ground-state energy increases
from the minimum to the maximum value, the corresponding
concurrence reduces from the maximum value to the minimum. The
minima of the concurrence also correspond to the energy level
crossings of the ground state and the first excited state.
\begin{figure}
\includegraphics[width=7cm]{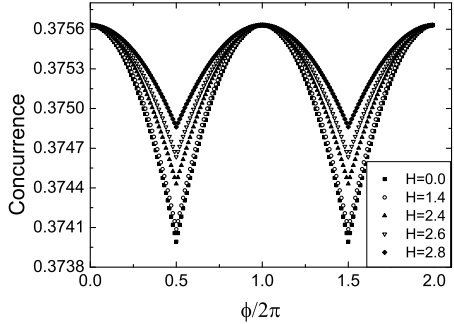} \caption{\label{Figure H-C}
The concurrence as a function of flux at different magnetic fields
when $\Delta=0.5$. The lines with $h\neq0$ are shifted along $y$
axis in order to make the maximal value be the same.}
\end{figure}
In the presence of the external magnetic field, the ground state
in the same fashion as Fig.(\ref{Figure ground}). We calculate the
ground state concurrence as a function of flux at different
magnetic fields, and find that when the magnetic field increases,
the value of the concurrence also increases. However, the
amplitude change of the concurrence decreases as the magnetic
field increases as we can see in Fig.(\ref{Figure H-C}), after we
shift these different lines along $y$ axis so that the maximum
coincide. When the spins are all fully polarized, the concurrence
is independent of the flux.

\section{SUMMARY and discussion}
In summary, we have studied the properties of the ground state
persistent current and the entanglement defined by concurrence in
the anisotropic Heisenberg ring pierced by a magnetic flux $\phi$.
By solving the BAE numerically, we obtain the ground state energy,
the persistent current, and the concurrence, which are all
periodic functions of the magnetic flux. The increasing anisotropy
$\Delta$ reduces the amplitude of the persistent current. The
locations of minimum concurrence are those of maximum persistent
current where an energy level crossing occurs. In the presence of
an external magnetic field, we find that when the value of the
magnetic field increased, the amplitude of the persistent current
decreases. When the magnetic field is larger than a certain value
($h=3.0$ for $\Delta=0.5$, when the spins are fully polarized),
the persistent current vanishes, and the concurrence is unchanged.
The value of concurrence also increases as the magnetic field
increased, but the amplitude change of the concurrence decreases.
\section{ACKNOWLEDGEMENTS}
We would like to thank Prof. X. Wan a critical reading of the
manuscript, and thank Prof. X. G. Wang some useful advices. This
work was supported by NSFC grant No.10225419.

\end{document}